\documentstyle[prl,aps,multicol,epsfig]{revtex}
\input{epsf}
\begin{document}
\draft
\title{Controlling the energy flow in nonlinear lattices:
a model for a thermal rectifier.}  
\author{M. Terraneo$^{1,3}$, M. Peyrard$^2$ and G. Casati$^{1,3,4}$
}  
\address{
$^1$ International Center for  the Study of Dynamical Systems,\\  
Universit\'a degli Studi dell'Insubria,
Via Valleggio 11,
22100 Como, Italy
\\
$^2$ Laboratoire de Physique, Ecole Normale Sup\'erieure de Lyon, 69364
Lyon cedex 07, France\\
$^3$ Istituto Nazionale per la Fisica della Materia, Unit\'a di Como, Via 
Valleggio 11, 22100 Como, Italy\\
$^4$ Istituto Nazionale di Fisica Nucleare, Sezione di Milano, Via Celoria 
16, 20133 Milano, Italy
} 
\address{
}
\address{
}   
\date{\today}
\maketitle

\begin{abstract}
We address the problem of heat conduction in 1-D nonlinear chains; we
show that, acting on the parameter which controls the strength of the
on site potential inside a segment of the chain, we induce a
transition from conducting to insulating behavior in the whole
system.  Quite remarkably, the same transition can be observed by
increasing the temperatures of the thermal baths at
both ends of the chain by the same amount. 
The control of heat conduction by nonlinearity
opens the possibility to propose new devices such as a thermal rectifier.    
\end{abstract}

\begin{multicols}{2}
\narrowtext

In recent years a renewed attention has been directed to the energy
transport in dynamical systems, a problem which has been denoted by
Peierls as one of the outstanding unsolved problems of modern physics
\cite{Peie}. These efforts mainly focused on the possibility to derive
the Fourier law of heat conduction on purely dynamical grounds without
recourse to any statistical assumption
\cite{Bon,Ford,Prosen,Lep,Hata,Camp,Giardina,Gendelman,Hu1,LIVI}. 
In spite of relevant
progresses, several problems remain open and we are still far from a
complete understanding\cite{Lep,Hata,Camp,Giardina,Gendelman,Hu1,LIVI}.

In this paper we investigate a 
different and important problem namely the possibility to
control the energy transport inside a nonlinear 1D chain
connecting two thermostats at different temperatures. 
We show that we can parametrically
control the heat flux through the system by acting on a small central part of
the chain. Even more interestingly, we show that it is possible to 
adjust the heat flux by varying the temperatures of 
both thermostats, keeping constant the temperature difference.
Thus we  provide a simple mechanism to change the properties of the 
system,  from a normal conductor obeying Fourier
law, down to an almost perfect insulator. 
Controlling  heat conduction by nonlinearity opens new
possibilities, such as the design of a lattice that carries heats
preferentially in one direction, i.e.\ a thermal rectifier.

We consider the  Hamiltonian 
\begin{equation}
H = \sum_{n=1,N} \frac{p_n^2}{2 m} + V_n(y_n) + \frac{1}{2} K 
(y_n-y_{n-1})^2
\label{PBmodel}
\end{equation}
\noindent
which describes a chain of $N$ particles with harmonic coupling of
constant
$K$ and  a Morse on-site potential $V_n(y_n)=D_n(e^{-\alpha_n  y_n}-1)^2$. 
This model was introduced for DNA chains where $m$ is
the reduced mass of a base pair, 
$y_n$ denotes the stretching from equilibrium position  of the hydrogen 
bonds connecting the two bases 
of the $n$-th pair and $p_n$ is its momentum \cite{pey1,Pey2,pey3}. 
In the context of the present study, model (\ref{PBmodel}) can simply
be viewed  as a generic system of anharmonic coupled oscillators, the
on-site potential arising from interactions with other parts of the
system, not included in the model. The Morse
potential is simply an example of a highly anharmonic soft potential,
which has a frequency that decreases drastically when the amplitude of
the motion increases.

In this paper we consider the out-of-equilibrium properties of
model (\ref{PBmodel}) by
numerically simulating the dynamics of the $N$ particle chain,
coupled, at the two ends,  with thermal baths at different temperatures 
$T_1$  and $T_2$.
We thermalize at $T_1$ and $T_2$ the first and the last $L$ particles
by using  
Nos\'e-Hoover thermostats chains  \cite{Nose,Marty}, 
or a Langevin description when we
investigate cases very far from equilibrium. 
We then integrate the differential equations of motion both for the  
thermostats and for the chain with a 4th order Runge-Kutta method as 
described in \cite{Runge}. The baths temperatures $T_1$, $T_2$ are
never large enough to drive the system beyond the thermal denaturation
temperature $T_c$ above which the mean value of $y_n$ diverges \cite{pey3}.

We compute  the temperature profile inside the system, i.e.\
the local temperature at site $n$ defined as  
$T_n = m \langle \dot{y_n}^2 \rangle $,   
where $\langle \hspace{0.2cm}
\rangle$ stands for temporal average, and the local heat flux 
$J_n 
= K  \langle \dot{y_n} (y_n - y_{n-1})\rangle$ \cite{Hu}.
The simulations are performed long enough to allow the system to reach a
non-equilibrium steady state with local thermal equilibrium where 
the local  heat flux is constant along the chain. 

\begin{figure}
\centerline{\epsfxsize=6.7cm\epsffile{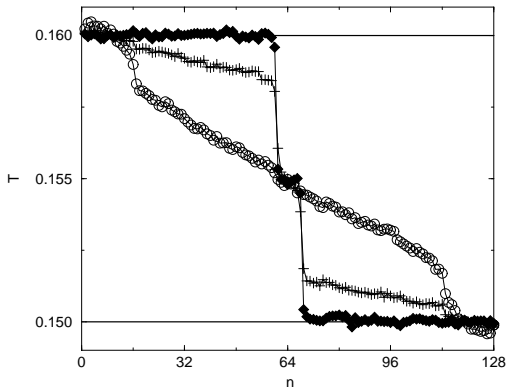}}
\caption{Temperature profile  of the chain with parameters  
$D=0.5$, $N=128$, $\alpha=1$, $K=0.3$, $M=8$, and 
$T_1=0.16$, $T_2=0.15$. The Morse potential constants are $D_1=D=0.5$  
 - homogeneous system- (circles), $D_1=1.2$ (diamonds) and 
$D_1=0.8$  (+)}
\label{fig1}
\end{figure}

As a preliminary step we have  considered the homogeneous case in 
which 
$D_n = D$, $\alpha_n=\alpha$, $n=1,..N$. Here, as expected, we have   
detected a 
temperature gradient inside the chain, 
as shown in Fig.~\ref{fig1} (circles), 
and we have verified that the system obeys the Fourier law of heat 
conduction. For fixed temperatures $T_1$ and $T_2$, and a chain length
varying from $N=64$ to $N=1024$ the heat flux evolves as $1/N$ as
expected for a system with a well defined, finite, thermal conductivity.

\begin{figure}
\centerline{\epsfxsize=6.7cm\epsffile{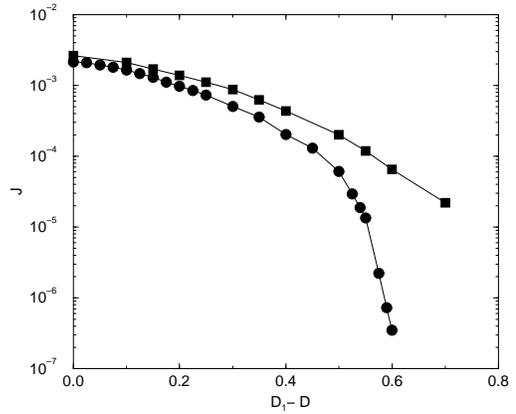}}
\caption{Heat flux as a function  of parameter $D_1 - D$; the squares 
refer to the Morse potential, while the circles refer to the corresponding 
harmonic limit. Here $N=128$, $D=0.5$, $ K=0.3$, $\alpha=1$, $L=16$}
\label{fig2}
\end{figure}

We now divide the chain between the thermostats in three regions in
which $D_n$ takes different values. In the left and right regions,
$n=1,\ldots (N - M)/2$ and $n=(N+M)/2+1, \ldots, N$, respectively, we
set $D_n = D$, while in the M sites of the central region $D_n=D_1$;
$\alpha_n = \alpha$ for the whole chain.  Typically, in our
simulations, we take $D=0.5$, $\alpha =1$, $K=0.3$, $m=1$, $N=128$,
$L=16$ and $M=8$ while $D_1$ has been varied from $0.5 $ to $1.2$.  As
shown by the numerically computed temperature profiles of
Fig.~\ref{fig1}, the results are now strikingly different.  The
temperature profile changes as $D_1-D $ is increased until, for large
enough $D_1$, the central region behaves as an insulator, the left and
right regions being thermalized at $T_1$ and $T_2$ respectively, and
the heat flux drops to negligible values $\sim 10^{-5}$.
Therefore the 
change in $D_1$ can induce a conductor--insulator transition
confirmed by
Fig.~\ref{fig2} which shows that the averaged heat flux $J$
decreases by two orders of magnitude when $D_1-D $ increases
from $0$ to $0.7$.

In order to understand this phenomenon, let us 
consider a linearized version 
of model (1), which is obtained by linearizing the Morse potential around 
$y_n=0$. The Hamiltonian now writes:

\begin{equation}
H = \sum_{i=n,N} \frac{p_n^2}{2 m} +\tilde{D_n} y_n^2 
+  \frac{1}{2} K  (y_n-y_{n-1})^2
\label{harmy}
 \end{equation}
where $\tilde{D_n}=D_n \alpha_n^2$ and $D_n$, $\alpha_n$  are defined above.

Numerical simulations with Hamiltonian (\ref{harmy}), show that 
this system displays the same conductor-insulator behavior in a
clearer and sharper way. Indeed, due to
integrability, there are some typical pathologies, such as the presence 
of plateaus instead of gradients 
in the temperature profile, as predicted by  \cite{Lebo}. 
This is why the transition from conducting to insulating 
regime  is even sharper than in the nonlinear
case (Fig.~\ref{fig2} (circles)~) until a cut-off is reached.

The  equations of motion for the system (\ref{harmy}), without thermostats,   
have plane wave solutions, 
$y_n(t)=e^{i 
k n - i \omega t}$, where the frequency $\omega$ and momentum 
$k$ satisfy the dispersion relation 
$\omega^2 = 2 K +2 \tilde{D} -2 K \cos{k} $
so that the allowed frequencies  are included in 
the so called \emph{phonon band} $2 \tilde{D} \le 
\omega^2 \le 2 \tilde{D} +4 K $.

Let us  now consider the harmonic and inhomogeneous chain, with 
$\tilde{D_1}  \ne \tilde{D}$ in  the central region.
In order to propagate through the system a wave should
satisfy both dispersion relations  $\omega^2 = 2 K 
+2 \tilde{D} -2 K \cos{k} $ and $\omega^2 = 2 K +2 \tilde{D_1} -2 K 
\cos{k_1} $.
If $\tilde{D_1} \ne \tilde{D}$ only a fraction of the allowed frequencies
of the waves coming from the left or right regions can 
satisfy both dispersion equations with real $k_1$, and therefore 
also propagate  
through the central region. For the remaining frequencies, the second 
equation  can only 
be solved by imaginary $k_1 = i\tilde{k_1}$ which gives rise to states   
$y_n \sim 
e^{-\tilde{k_1} n}$ which decrease exponentially from the edges of the central 
region and therefore cannot propagate through it so that their contribution to 
the total heat flux $J$ is exponentially small. 
 By increasing the  separation of the two phonon 
bands  $\tilde{D_1}-\tilde{D}$, 
the fraction of frequencies which can give rise to plane 
waves inside the entire system falls down and reaches zero as 
$\tilde{D_1}-\tilde{D} > 2 K $, where the band separation 
is larger than the band width, and the 
bands do not overlap anymore. In this situation, waves coming from the  
thermostats are reflected back from the central region, the heat 
flux is exponentially small and the system acts as an almost perfect insulator.
In the case of partially overlapping bands one obviously expects an  
intermediate situation. The dependence of the heat flux on the band  
overlapping is shown by the lower curve in Fig.~\ref{fig2} and its
variation can be calculated with the above analysis.

The addition of a nonlinear term to the Hamiltonian does not change the  
substance of the above argument.
The numerically observed smoother decay of the heat flux vs. $D_1 -D $  
given by the upper curve in Fig.~\ref{fig2} 
is due to the fact that plane waves are not exact solutions 
for the nonlinear case and the presence of solitary excitations like 
 breathers may play an important role.

It is now interesting to investigate whether one can control the transport  
properties of the chain by acting on the temperature of the heat baths 
without  modifying the parameters of the interaction or of the on 
site potential.  
To this end we recall that, at constant $T$,  the nonlinear model
can be approximated by a fully harmonic Hamiltonian, 
with temperature dependent parameters $\Omega^2(T)$, $\Phi(T)$ \cite{Pey2}
\begin{equation}
H_0 = \sum_{n=1,N} \frac{p_n^2}{2 m} +\Omega^2(T) y_n^2 + \frac{1}{2} \Phi(T)  
(y_n-y_{n-1})^2
\end{equation}

\begin{figure}
\epsfig{file=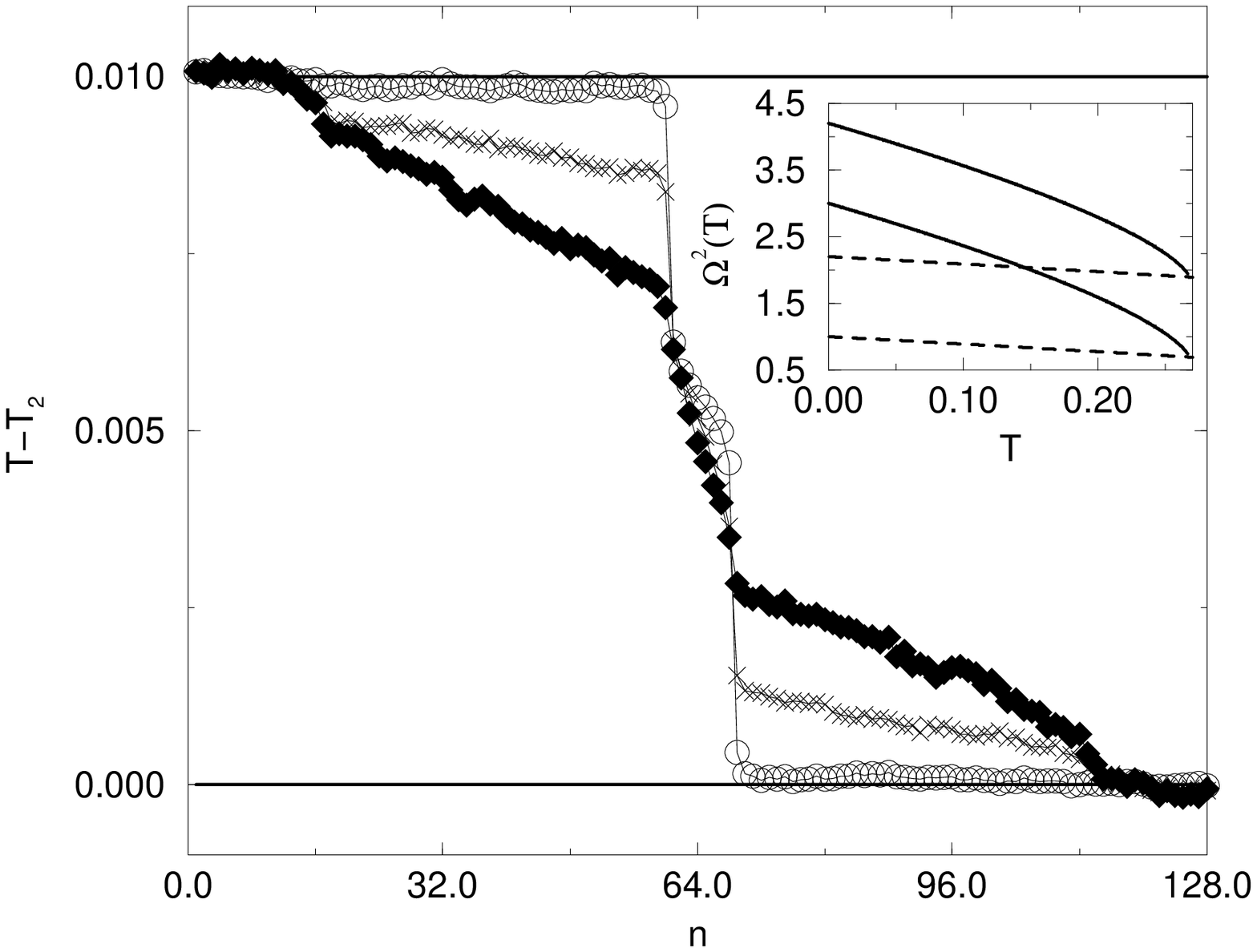,width=6.7cm}
\caption{Temperature profiles  for system 
(\ref{PBmodel}) at three different values of $T_2$ for the same
temperature gradient $T_2 - T_1$. Parameters are
 $N=128$, $M=8$, $\alpha_1=2$, $\alpha=1$, 
$D_1=0.375$,
$D=0.5$;  $T_2 =  0.09$ ( circles), $T_2=0.19$ (x) and 
$T_2=0.29$
(diamonds).\\
Inset: Phonon band profiles vs. temperature; 
the continuous lines represent the 
phonon  band for  $\alpha=2$,
$D=0.375$; the dashed lines  $D=0.5$, $\alpha=1$. The full phonon band is
shown with bandwidth  $4 K =1.2$}
\label{fig5}
\end{figure}

An effective-phonon analysis, shows that, 
while $\Phi(T) =K $ is temperature independent, $\Omega(T)$ decreases as 
$T$  increases \cite{Pey2}.
The lowering of the band vs. the temperature is of course determined
by the parameters of the system, $\alpha_n$, $D_n$.  Typical band
profiles are shown in the inset of Fig.~\ref{fig5}, for different
parameter choices.

We may now choose parameters $\alpha=1$, $D=0.5$ (dashed lines in 
the inset of Fig.~(\ref{fig5})) for the left
and right regions, and  $\alpha_1=2$, $D_1=0.375$ (full line  
in the inset of Fig.~(\ref{fig5})) for the central region. 
From  the inset of Fig.~(\ref{fig5})
it is clearly seen that, at low $T$, the phonon  band of the central 
region does not
overlap with the band of the other two side regions.
 In this situation, the heat flux is nearly zero and the temperature 
profile  shows a clear  insulating behavior, with first region 
thermalized at  temperature $T_1$ and third region at temperature $T_2$ as 
shown in  Fig.~\ref{fig5} (circles).
As we increase  both  temperatures $T_1$ and $T_2$, 
keeping $|T_1-T_2|$
 constant, the bands start to overlap, the temperature profile exhibits a 
gradient and the heat flux increases drastically (Fig.~\ref{fig5}). 

As temperature increases,  nonlinearity becomes more  and more 
important  and the effective phonon
analysis becomes less and less meaningful. Nevertheless, numerical analysis
 shows that  the above argument  of band 
overlap still remains valid  even if a clean thermal
gradient is not obtained and therefore the system does not
become a perfect conductor. 

\begin{figure}
\centerline{\epsfig{figure=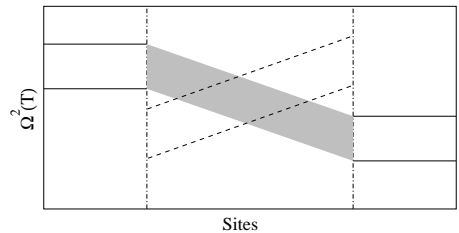,height=3.0cm}\bigskip}
\caption{Schematic picture of the phonon bands in a ``thermal
rectifier'' for two directions of the temperature gradient. The bands in
the left and right weakly anharmonic regions do not change
significantly with the
orientation of the gradient. In the central part, when the high
temperature side is on the right, the band evolves in space as shown
by the shaded region, while, if the high temperature side is on the
left, the band evolves according to the dashed lines.}
\label{figdiode}
\end{figure}

Once the general mechanism of the thermal conduction in a composite
nonlinear lattice has been understood, this opens many
possibilities. For instance one can design a {\em thermal rectifier}, as
schematized on Fig.~\ref{figdiode}. 
A strongly nonlinear region is
sandwiched between two weakly anharmonic left and right domains. In
the presence of a thermal gradient in the central part, the 
effective phonon
frequencies evolve in space in a way that depends on the orientation
of the gradient. This can provide either a good matching of the bands
at the interfaces, leading to a thermal conduction across the system, 
or a complete mismatch leading to poor conduction. This picture is
oversimplified for various reasons: the self-consistent phonon
method is approximate, there is not a homogeneous gradient in the
system due imperfect thermal contact at the interfaces, and, in spite
of nonlinear mixing, in the conducting direction, 
the frequencies injected at one side of the
central region do not shift gradually, following the phonon band, to
match the band of the other outer region.

\begin{figure}
\centerline{\epsfig{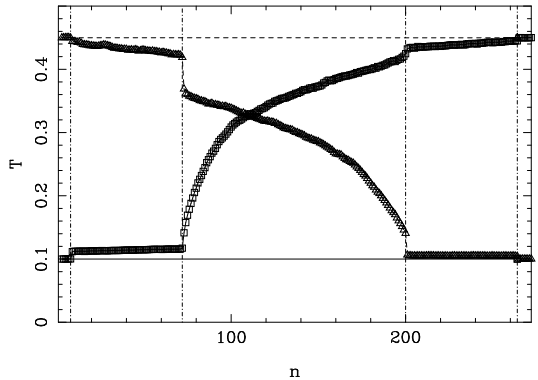}\bigskip}
\caption{Temperature profiles in a ``thermal rectifier'' for two opposite
orientations of the thermal gradient. The dash-dotted lines show the
borders of the different regions in the lattice. The thermostats have
a size $L=8$, the central region a size $M=128$, and the left and
right regions contain 64 particles. The coupling constant is $K=0.18$,
and the parameters of the Morse potential are $D=4.5$, $\alpha=0.5$,
$D_1=0.7$, $\alpha_1 = 1.4$, $D' = 2.8$, $\alpha' = 0.5$ in the left,
central and right regions respectively. The temperatures of the
thermostats are $T=0.1$ and $0.45$. When the high temperature is on the
right side of the lattice, the average flux is $J = 0.146~10^{-3}$ 
while, when the thermal gradient is reversed the flux drops to 
$J=0.755~10^{-4}$. Note in this case the discontinuities at the
interfaces between region which attest of the bad energy transfer at
these points.}
\label{fig:diode2}
\end{figure}

In spite of these restrictions Fig.~\ref{fig:diode2} shows that 
it is possible to build a simple thermal
rectifier  where the thermal flux changes by a factor of about 2
when the direction of the thermal gradient is reversed. In fact
thermal rectifiers are already known: a layer of fluid, heated from below or
from above yields very different thermal flows because the
former case leads to convection and the latter does not. To our
knowledge, it is nevertheless the first time that a ``solid system''
(a highly simplified model of a solid indeed) shows such a
behavior. 

In this work we have studied a specific nonlinear lattice but the
results rely on ideas of general validity, the matching of phonon
bands or the frequency shift associated to nonlinearity, so that one
can expect that various physical systems could exhibit similar
behaviors. The simple model that we have chosen to introduce these ideas
involves an on-site potential which describes the effect of an
external substrate not explicitely included in our calculations. In
an actual device, such a substrate would introduce a second channel
for thermal energy transport, operating in parallel with the channel
studied here. This could reduce the efficiency of the rectifier, unless
the thermal conductivity of the substrate is significantly lower than
the conductivity of the lattice of nonlinear oscillators. Any solid
thermal insulator or weak conductor, can, in principle, be used as a
substrate. A prototype could be a disordered harmonic lattice
\cite{LIVI}.

Among the possible applications of this study, one can think of
biological molecules.  Controlling the energy flow in biomolecules is
important, for instance when the energy of ATP hydrolysis is released
locally in a molecular motor to be used elsewhere after some
delay\cite{Yanagida}. 
Electronic states could be involved, but we show here that a
control of the flow of thermal energy can also be achieved through the
strong nonlinearities which exist in biological molecules and could
even be tuned by events such as a conformational change. Experiments
show that the energy flow in proteins can be extremely slow
\cite{Champion}, providing a weakly conducting substrate. On the other
hand our results show that a controlable conducting channel attached
to this substrate is conceivable. Whether biomolecules exploit this
possibility is still a completely open problem.

M.T. thanks the Laboratoire de Physique of ENS Lyon, where part of
this work has been performed, for its hospitality and computing
facilities.
M.P. would like to thank B. Castaing (ENS Lyon) for helpful
discussions.
This work has been partially supported by EU Contract
No. HPRN-CT-1999-00163 (LOCNET network) and by MURST (Prin 2000, 
\emph{Caos e localizzazione in Meccanica Classica e Quantistica}).

\end{multicols}

\end{document}